\documentclass{caosp309}

\usepackage{graphicx}

\usepackage{natbib}
\usepackage[normalem]{ulem}
\articleNo{301}
\pubyear{2023}
\volume{53}
\volnumber{4}
\firstpage{175}
\received{July 20, 2023}
\accepted{September 14, 2023}

\def\BibTeX{{\rm B\kern-.05em{\sc i\kern-.025em b}\kern-.08em
             T\kern-.1667em\lower.7ex\hbox{E}\kern-.125emX}}

\begin{document}

%
\htitle{Timing of accreting neutron stars with future X-ray instruments\ldots}
\hauthor{Vladim\'{\i}r Karas et al.}

\title{Timing of accreting neutron stars with future X-ray instruments: towards new constraints on dense matter equation of state}

%
%
\author{
        V.~Karas\inst{1}\orcid{0000-0002-5760-0459}
        \and
        K.~Klimovi\v{c}ov\'{a}\inst{2}\orcid{0000-0002-0930-0961}
        \and
        D.~Lan\v{c}ov\'a\inst{2}\orcid{0000-0003-0826-9787}
        \and 
        M.~\v{S}tolc\inst{1}\orcid{0009-0001-7020-2711}
        \and  
        J.\,Svoboda\inst{1}\orcid{0000-0003-2931-0742}
        \and
        G.~T\"or\"ok\inst{2}\orcid{0000-0003-3958-9441}
        \and
        M.~Matuszkov\'{a}\inst{2}\orcid{0000-0002-4193-653X}
        \and
        E.~\v{S}r\'{a}mkov\'{a}\inst{2}\orcid{0009-0000-7736-6180}
        \and
        R.~\v{S}pr\v{n}a\inst{2}\orcid{0000-0002-5085-9740}
        \and
        M.~Urbanec\inst{2}\orcid{0000-0001-9635-5495}
       }

%
\institute{
           Astronomical Institute of the Czech Academy of Sciences\\ Bo\v{c}n\'{\i} II 1401, CZ-14100, Prague, Czech Republic\\ \email{vladimir.karas@asu.cas.cz}
         \and 
           Research Centre for Computational Physics and Data Processing,\\ Institute of Physics, Silesian University in Opava, Bezru\v{c}ovo n\'am.\ 13,\\ CZ-746-01 Opava, Czech Republic\\
          }

\date{March 8, 2003}

\maketitle

\begin{abstract}
The Enhanced X-ray Timing and Polarimetry (eXTP) mission is a space mission to be launched in the late 2020s that is currently in development led by China in international collaboration with European partners. Here we provide a progress report on the Czech contribution to the eXTP science. We report on our simulation results performed in Opava (Institute of Physics of the Silesian University in Opava) and Prague (Astronomical Institute of the Czech Academy of Sciences), where the advanced timing capabilities of the satellite have been assessed for bright X-ray binaries that contain an accreting neutron star (NS) and exhibit the quasi-periodic oscillations. 

Measurements of X-ray variability originating in oscillations of fluid in the innermost parts of the accretion region determined by general relativity, such as the radial or Lense-Thirring precession,  can serve for sensitive tests enabling us to distinguish between the signatures of different viable dense matter equations of state. We have developed formulae describing non-geodesic oscillations of accreted fluid and their simplified practical forms that allow for an expeditious application of the universal relations determining the NS properties. These relations, along with our software tools for studying the propagation of light in strong gravity and neutron star models, can be used for precise modeling of the X-ray variability while focusing on properties of the intended Large Area Detector (LAD). 

We update the status of our program and set up an electronic repository that will provide simulation results and gradual updates as the mission specifications progress toward their final formulation.
\keywords{stars: black holes -- stars: neutron -- X-rays: binaries -- polarimetry}
\end{abstract}

%
\section{Introduction}
\label{intr}
The Enhanced X-ray Timing and Polarimetry (eXTP) project is an international project led by China with European contribution \citep{2019SCPMA..6229502Z}. eXTP will be a science mission designed to study the state of matter under extreme conditions of density, gravity, and magnetism \citep{2019SCPMA..6229506I,2019SCPMA..6229503W,2019SCPMA..6229505S}. It aims to address three scientific questions in fundamental physics.

Among the primary goals are the long-standing problems of determining the Equation of State (EoS) of matter at supra-nuclear density, the study of accretion in the strong-field regime of gravity, and the measurement of quantum-electrodynamic (QED) effects in highly magnetized stars \citep{1971NuPhA.175..225B,2019BAAS...51c.362U}. Primary targets include compact objects: isolated NS as well as those in binary systems, cosmic objects with strong magnetic fields such as magnetars, and stellar-mass, supermassive and tentative intermediate-mass black holes \citep{2019SCPMA..6229504D}. eXTP will provide a significant enhancement of sensitivity and scheduling flexibility of the parametric measurements, which will build on the exciting results of the currently active IXPE \citep{2023NatAs...7..635W,2021AJ....162..208S}. Indeed,
the magnetic intensity in these objects can reach almost a billion times higher magnitude than what is achievable on Earth.

Almost exactly a decade ago at this conference (10th INTEGRAL/BART Workshop in Karlovy Vary, 22-25 April 2013), we explored the appearance of X-ray signal generated by hot spots moving along quasi-elliptic trajectories close to the innermost stable circular orbit in the Schwarzschild spacetime. The aim of our investigation was to reveal whether the observable characteristics of the Fourier power-spectral density can distinguish between the competing models \citep{2014AcPol..54..191K}. In the present contribution, we elaborate on the approach of X-ray lightcurve timing analysis to constrain the NS EoS. In particular, we are interested in the prospects of the upcoming space-borne technology that has been meanwhile developed.

eXTP is an enhanced proposal based on its predecessor, the XTP mission concept, that has been selected and funded as one of the so-called background missions in the Strategic Priority Space Science Program of the Chinese Academy of Sciences since 2011. The strong European participation has significantly enhanced the scientific capabilities of eXTP \citep{2022SPIE12181E..1XF}. Likewise, the Italy-led development of the Large Area Detector is based on the prior LOFT proposal for a mission of the European Space Agency \citep{2012ExA....34..415F,2016SPIE.9905E..1RF,2001Natur.411..662C}. In January 2017, the Chinese National Space Agency accepted the eXTP mission as a flagship space science mission, initiated and led by China in an international collaboration \citep{2016SPIE.9905E..1QZ}. The planned launch date of the mission is in the late 2020s. The science observations are planned for at least eight years with the possibility of further extensions.

The mission carries a unique and unprecedented suite of state-of-the-art scientific instruments enabling for the first time ever the simultaneous spectral-timing-polarimetry studies of cosmic sources in the energy range from 0.5--30 keV (and beyond).

\begin{figure}[tbh!]
\centerline{\includegraphics[width=\textwidth]{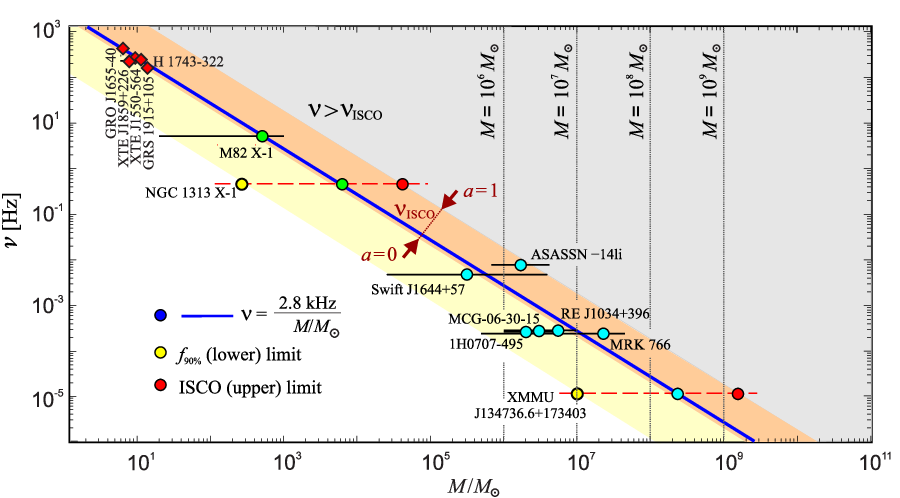}}
\caption{Quasi-Periodic Oscillations from accreting black holes across a wide range of mass. Enhanced sensitivity and a multitude of instruments will make eXTP an ideal mission to explore accretion flow oscillations from a variety of black holes over the range of mass. Whereas the upper left section of the plot corresponds to stellar-mass black holes in microquasars observed in the Milky Way, the lower right section is populated by supermassive black holes in cores of active galaxies. The orange stripe indicates the uncertainty range corresponding to the spin parameter in the range from a non-rotating black hole ($a=0$) up to maximum rotation ($a=1$). The yellow stripe gives the uncertainty corresponding to the luminosity distribution from the inner part of the standard (Keplerian) accretion disk, where more than 90\% of the entire disk luminosity is produced. 
}
\label{f1}
\end{figure}

In a brief contribution, we discuss promising scenarios regarding the detection and analysis of quasi-periodic phenomena in the context of eXTP, where its large collection area in the most relevant keV range and advanced timing capabilities offer a leap in our understanding of the puzzling sources. Our research collaboration at the Astronomical Institute of the Czech Academy of Sciences in Prague and the Institute of Physics of the Silesian University in Opava takes part in the work on data simulations and detector performances related to study of accreting NS and the electromagnetic signatures of the internal structure via high-frequency quasi-periodic oscillations (HF QPOs) \citep[e.g.][and further references cited therein]{1999ApJ...526..953K,2014MNRAS.439.1933B,2014AcPol..54..191K,2015A&A...578A..90S,2019MNRAS.488.3896T,2020A&A...643A..31K, 2022ApJ...929...28T}. 
An updated overview has been given in a recent Thesis \citep{DebiThesis}.

\section{Rapid variability of accreting black holes and neutron stars}

Accreting black holes (both stellar mass and supermassive) and NSs exhibit extraordinarily complex phenomenology of X-ray variability patterns on various timescales, including the shortest ones of millisecond periods \citep[and more references cited therein]{1998AdSpR..22..925V,2006csxs.book..157M,2016AN....337..398M}. Here we focus on the widely discussed twin kHz peaks that have been detected in the Fourier power spectra of a few dozen of X-ray binaries, and also in several AGN and the Galactic centre supermassive black hole Sgr A* with correspondingly down-scaled frequency \citep{2003PASJ...55..467A,2004ApJ...617L..45B,2005AN....326..856T,2019A&A...622L...8G}. The position of the two peaks, and of the third (burst oscillation) one, as well as their significant coherence, limit the variety of potential models of the geometry of the system. 
Figure~\ref{f1} shows the scaling relation of selected HF QPO sources in the QPO frequency vs.\ black hole mass diagram; see also a detailed caption in recent updates and analyses in \citet{2021IAUS..356..348S,2023AN....34420114A}. The figure has been amended from the original work of \citet{2019A&A...622L...8G} by revising data points for several objects and including additional sources (XTE J1859+226, H 1743-322, ASASSN-14li).\footnote{References to individual sources: \cite{2001ApJ...554.1290G,2001ApJ...552L..49S,2002ApJ...568..845O,2002ApJ...580.1030R,2001Natur.414..522G,2004AAS...20510405R,2014Natur.513...74P,2015ApJ...811L..11P,2011ApJ...738L..13M,2012Sci...337..949R,2016A&A...594A.102C,2008Natur.455..369G,2016ApJ...830..136B,2018A&A...616L...6G,2016ApJ...832..197H,2003MNRAS.343..164B,2016ApJ...819L..19P,2006A&A...445...59T,2017ApJ...849....9Z,2018MNRAS.477.3178C,2022MNRAS.517.1476Y,2022MNRAS.517.1469M,2017MNRAS.471.1694W,2019Sci...363..531P,2017ApJ...834...88M,2015ApJ...798L...5Z}.}

Very recently, a novel phenomenon of quasi-periodic eruptions (QPE) and outflows (QPOouts) has been diagnosed in a growing number of nuclei of external galaxies \citep{2019Sci...363..531P}; the latter events have been tentatively associated with Tidal Disruption Events (TDE) \citep{2021ARA&A..59...21G}. The effects of general relativity cannot be ignored in modeling systems where the signal is emerging from the immediate vicinity of the inner accretion region near a compact body \citep{1999PhRvL..82...17S,1999ApJ...524L..63S,2019NewAR..8501524I}. It has been argued that the oscillation frequencies can be employed to constrain the viable EoS of the NS matter \citep{2007PhR...442..109L} and to test the TDE mechanism and even the validity of the no-hair theorem in the case of black holes space-times.

\begin{figure}[tbh!]
\centerline{\includegraphics[width=\textwidth]{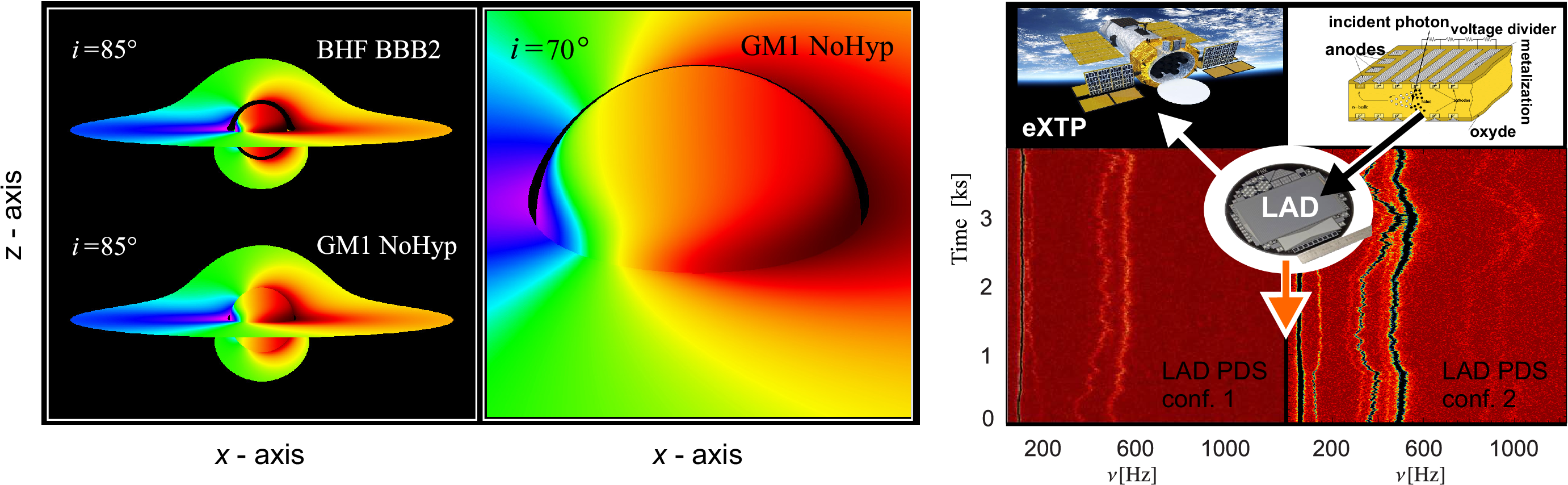}}
\caption{Left: Images illustrating Doppler shifts effects for Keplerian disks around a rotating NS calculated assuming the same NS mass and spin frequency but two different EoS. An emphasized view shows the spreading layer where matter from the disks enters the rotating NS surface. Illustration based on the work of \v{S}pr\v{n}a et al. (in preparation). Right: The LAD detector on board of the eXTP satellite (\url{isdc.unige.ch/extp}) and examples of simulated time evolution of the X-ray variability of NS accretion disk as predicted for the LAD detector 
\citep{2014MNRAS.439.1933B,2016SPIE.9905E..1QZ,2019SCPMA..6229504D,2022SPIE12181E..1XF}.
}
\label{figure:LAD}
\end{figure}

Here, we concentrate on predictions regarding suitable targets with respect to the low-mass X-ray binaries \citep[LMXBs;][]{2016ApJ...833..273T}, and we extend the range over the entire mass scale up to supermassive black holes. In the stellar close-binary systems, the mass is transferred from the companion star by overflowing the Roche lobe and forming an accretion disk or a torus that surrounds the NS. The accretion flow forms the main component that contributes to the high X-ray luminosity of these objects; most of the radiation arises from its inner region and the boundary layer adjacent to the surface. Oscillation properties of the flow should thus reflect the presence of the central body and they could help us to constrain its internal EoS. Particular attention has been attracted to the kilohertz QPOs. In the case of LMXBs, these oscillations occur in the range  of 100--1000~Hz; their intrinsic frequencies are thus comparable to the orbital timescale near the surface of the NS. This coincidence suggests that the emerging signal originates in the innermost region of the accretion flow close to the NS surface. As their name suggests, twin kHz QPOs are frequently observed in pairs; correlations between the two peaks can be used to test the viability of different models.

\section{Results}

\begin{figure}[tbh!]
\centerline{\includegraphics[width=\textwidth]{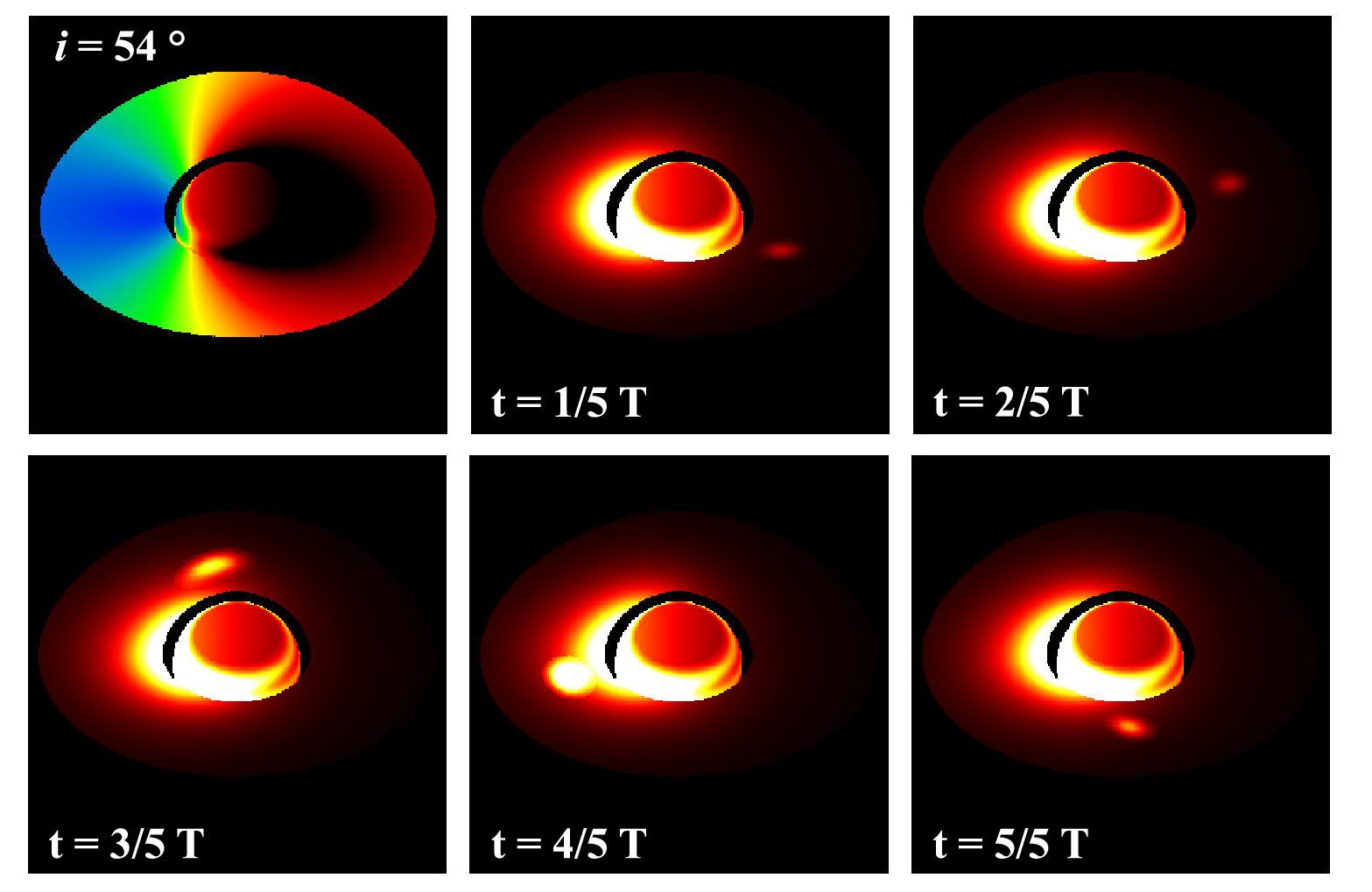}}
\caption{The technique of image mapping shows the Doppler shift effects, here in the case of a Keplerian disk containing an orbiting hot spot (blue-green-red color scale). Consecutive snapshots display the corresponding bolometric intensity of the observed radiation in different parts of the orbital period (red-to-yellow color scale).}
\label{figure:spot}
\end{figure}

We have developed a set of flexible simulation tools that are aimed at exploring suitable observational strategies and identifying the characteristics of promising targets for upcoming space programs. 
The focus has been set on recurrent quasi-periodic events. These allow for an expeditious application of various physical processes and searching for their observable signatures using the LAD detector; see Figure~\ref{figure:LAD} for an illustration of various aspects of the adopted methodology. Efficient ray-tracing computations with all General Relativity effects taken into account are essential to account for the flux variations as well as spectral, timing, and temporal characteristics \citep{1992MNRAS.259..569K,2004ApJS..153..205D,2005ragt.meet...11B,2008MNRAS.384..361D,2014MNRAS.439.1933B,2015A&A...581A..35B,2014AcPol..54..191K}.
Hereafter, we show the setup and the main parameters that govern our simulations.\footnote{Our simulation results are available together with a more detailed description in an electronic form at \url{astrocomp.physics.cz/simulator/}. Access to the data is open to all researchers; we ask for this article to be acknowledged if the data are used in any follow-up publications.}

\begin{figure}[tbh!]
\centerline{\includegraphics[width=0.95\textwidth]{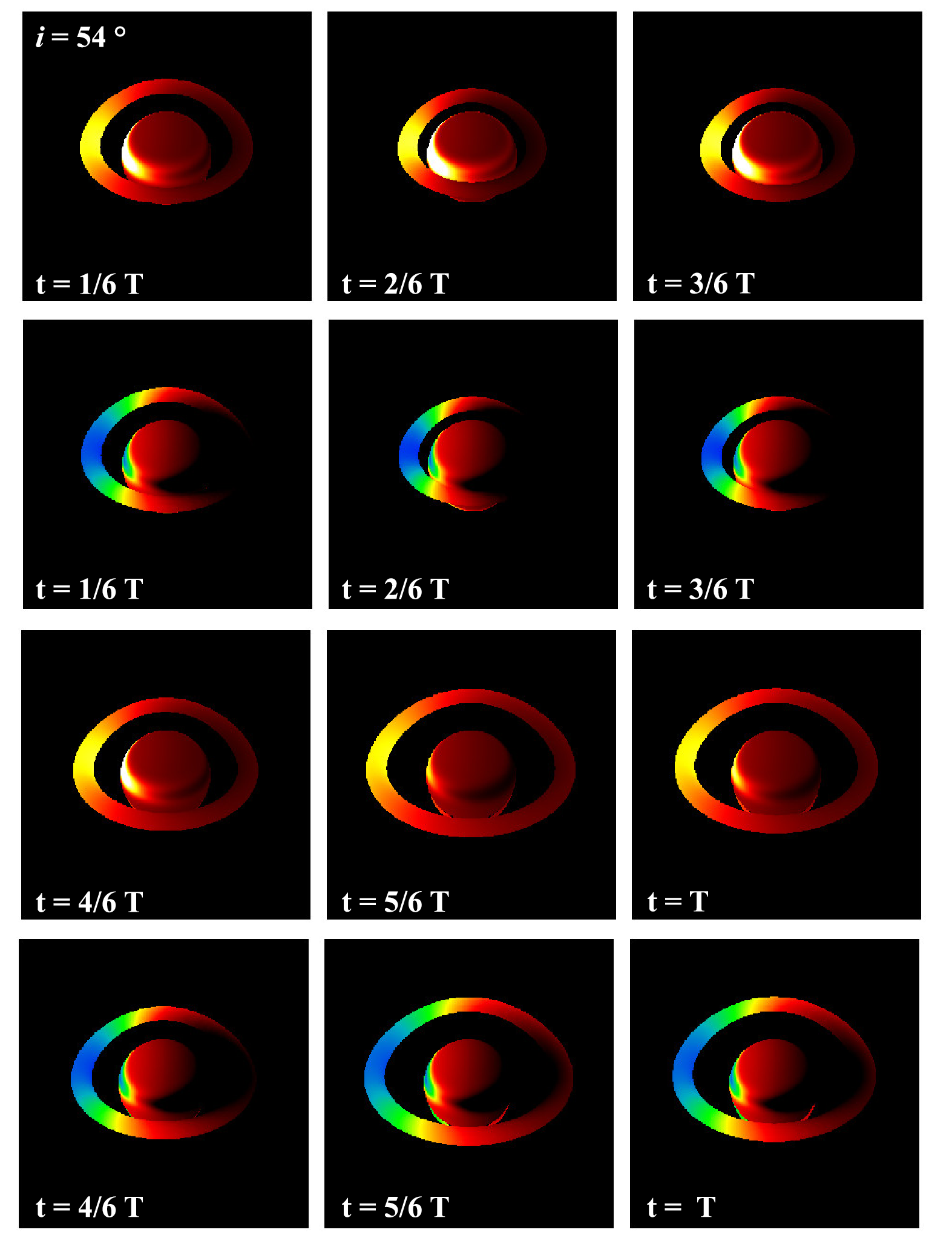}}
\caption{In analogy with the preceding figure, we show snapshots that map the Doppler energy shift effects for axisymmetric radial oscillations of an inner accretion torus (blue-green-red color scale) and the corresponding snapshots displaying the bolometric intensity of the observed radiation in different parts of the disk oscillation period (red-to-yellow color scale).}
\label{f3}
\end{figure}

\begin{figure}[tbh!]
\centerline{\includegraphics[width=0.95\textwidth]{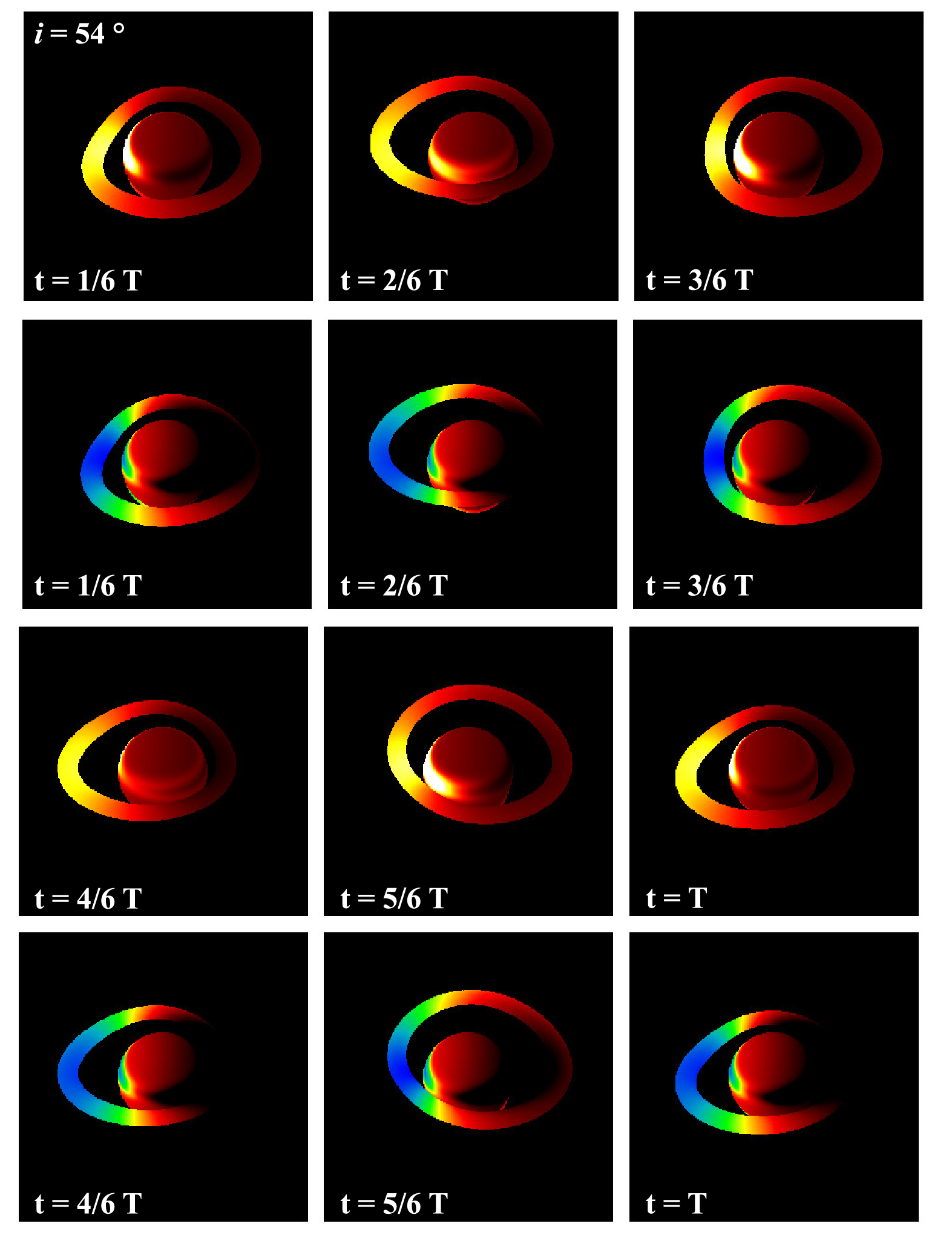}}
\caption{{A sequence of snapshots mapping Doppler shift effects in the case of the radial precession (non-axisymmetric oscillations) of an inner accretion torus (blue-green-red scale) and the corresponding snapshots showing the bolometric intensity of the observed radiation in different parts of the disk oscillation period (red-to-yellow scale).}}
\label{f4}
\end{figure}

\begin{figure}[tbh!]
\centerline{\includegraphics[width=0.95\textwidth]{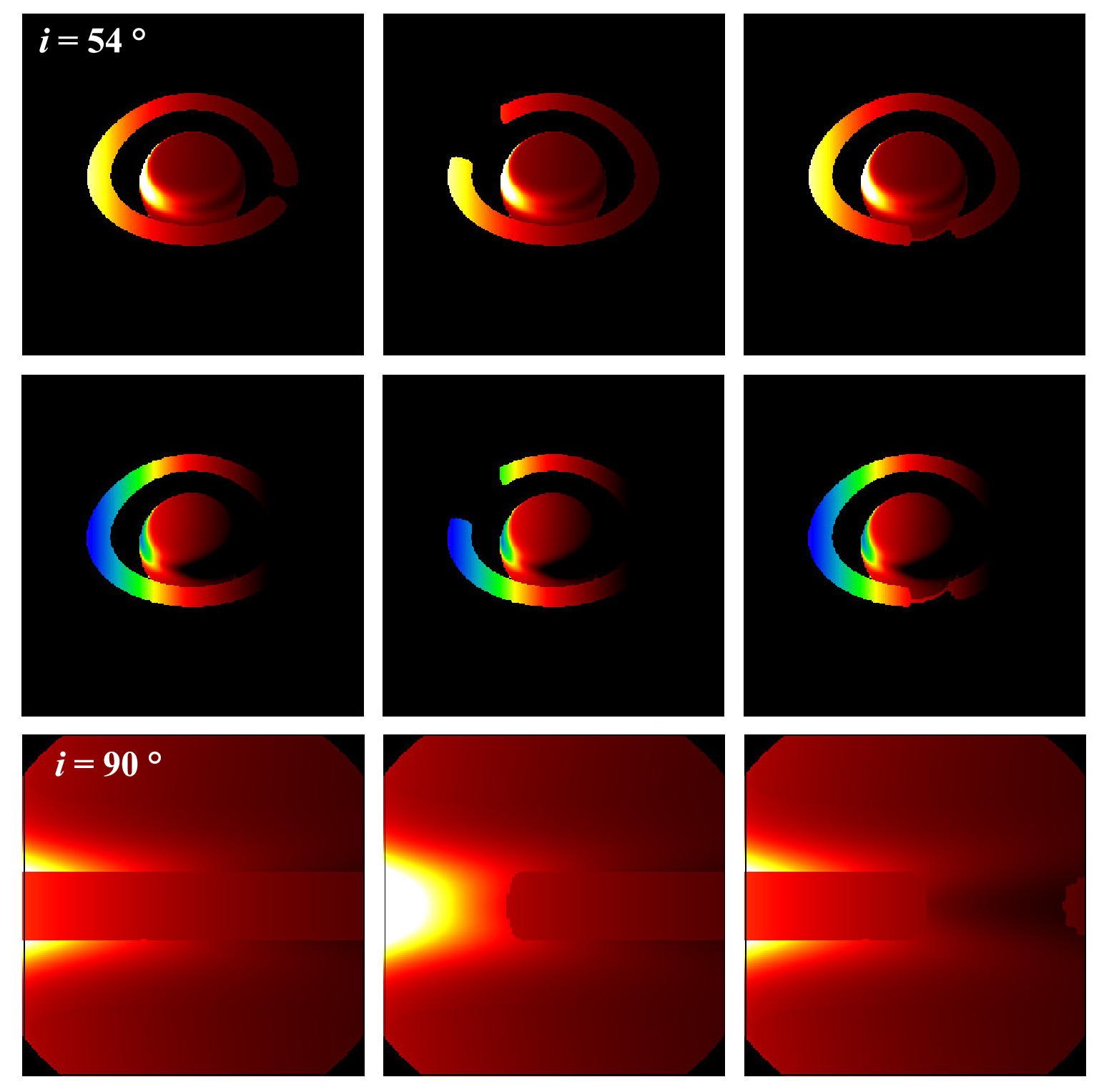}}
\caption{The snapshots displaying the bolometric intensity of the observed radiation in three different parts of the Keplerian period of a rotating product of torus instability (red-to-yellow color scale) along with the corresponding snapshots mapping the Doppler energy shift effects. The bottom row contains three images calculated for a near-equatorial view and very closely following parts of the Keplerian period, illustrating the strong variability of such a configuration. General Relativity effects are all taken into account in the ray-tracing computations.}
\label{figure:train}
\end{figure}

\subsection{Frequencies of QPOs: model predictions}

Comparison between the observed and expected QPO frequencies can help reveal the NS mass and angular momentum  \citep{2010AcA....60..149U,2022ApJ...929...28T}. To this end, sufficient sensitivity in the kiloelectronvolt band, as well as fine temporal and spectral resolution, are required. In \citet{2012ApJ...760..138T,2016ApJ...833..273T}, we studied within the framework of general relativity the mass--angular-momentum relations implied by a number of QPO models. 

In the above-quoted papers, we showed that the degeneracy between the mass and angular momentum constraints can be removed when the NS spin frequency is determined in an independent way. Furthermore, in \citet{2016ApJ...833..273T,2022ApJ...929...28T}, we extended our results to a large set of EoS and explored effects related to the NS quadrupole moment and its rotationally induced oblateness. Remarkably, we found that strong restrictions to the QPO model and the implied NS mass could be obtained when the low-frequency QPOs or X-ray burst measurements were taken into account.

\subsection{Examining full synthetic data, and beyond}

While calculations of the QPO frequencies enable comparisons between the observed and synthetic data offering fruitful outputs, a complete comparison of lightcurves that would include consideration of the QPO amplitudes, coherence times, and spectral properties is even more promising.

In the present contribution, we reported on an extended set of model simulations, where we take into account the expected specifications of eXTP detectors, namely, their collecting area and energy and timing resolution.

The current work highly benefits from using our software package, which is based on the consideration of NS models on the background of the Hartle-Thorne geometry \citep[\texttt{HTstar},][]{2013MNRAS.433.1903U,2019ApJ...877...66U} along with the key outputs of the LSD code core \citep[{\texttt{Lensing Simulation Device}},][] {2007CEJPh...5..599B,2014MNRAS.439.1933B,2015A&A...581A..35B}. Within the context of the QPO models, we assume various configurations of oscillating radiating accreted fluid (disks, tori, spots, spiral waves, etc.) along with a rotating NS that exhibits a hot spreading layer of enhanced X-ray emission. Example images of such configurations are given in Figures \ref{figure:spot}--\ref{figure:train}.

\section{Conclusions}

The enhanced X-ray Timing and Polarimetry (eXTP) mission has been developed by a consortium led by the Institute of High-Energy Physics of the Chinese Academy of Sciences and is envisaged to launch in the later-2020s. It is currently considered to be a priority of the upcoming astrophysics program. The Czech contribution is a part of the consortium of several European groups led by the Italian INAF (Istituto Nazionale di Astrofisica). eXTP will carry four instrument packages for the X-ray 0.5--50 keV bandpass, with the primary purpose to explore matter under extreme density, strong-gravity effects on spectra and polarization, and cosmic magnetism in and around compact objects. 

The scientific part of our contribution consists mainly of the performance analysis of the instrument to achieve the required scientific goals of the eXTP mission concerning various hardware configurations and different targets. In the present paper, we focused our attention on the dominant program of X-ray timing of accreting NSs. So far, commonly to several promising QPO models, even the QPO modulation mechanism
remains unexplained. Using our simulations we can show that a sufficiently strong modulation of the observed radiation arises when we consider modulation of accretion rate, which provides variable luminosity of the NS boundary layer and is combined with obscuration of the radiation by structures present in the innermost accretion region (see Figure~\ref{figure:train} for an illustration). 

QPOs offer an independent measurement of black hole mass and spin via modeling the low and high (kilohertz) frequency QPOs observed in the intermediate states of X-ray binaries. Low-frequency QPOs seen in the hard and intermediate states are thought to be produced by the Lense-Thirring precession of the inner hot flow. In this case, modeling the geometry of the precessing flow using tomography and polarimetry will give a direct measure of the offset between the black hole spin and disk angular momentum. This will allow us to explore the accretion flows down to ISCO. An analogy will be explored in the case of accreting supermassive black holes that exhibit the corresponding quasi-periodicities.

\medskip\noindent
\textbf{Data availability.} The synthetic data underlying this article will be shared upon reasonable request. Simulation results and tools will be further added to the above-cited repository (cf.\ footnote \#2), and they will be updated once new mission specifications become available.

\acknowledgements
Our research has been carried out in the context of Czech participation to the upcoming eXTP satellite mission. We acknowledge a continued support from European Space Agency PRODEX program titled ``Hardware contribution to the Chinese X-ray mission eXTP'' (ref.\ 4000132152), and from the Czech Science Foundation EXPRO project on ``Accreting black holes in the new era of X-ray polarimetry missions'' (ref.\ 21-06825X). We further acknowledge an internal grant of the Silesian University, $\mathrm{SGS/31/2023}$, and the Astronomical Institute's institutional support, RVO 67985815.

\bibliography{karas-ibws-2023}

\end{document}